# EXCELLENT PERFORMANCE OF 650 MHz SINGLE-CELL NIOBIUM CAVITY AFTER ELECTROPOLISHING*

V. Chouhan†, D. Bice, A. Cravatta, T. Khabiboulline, O. Melnychuk. A. Netepenko, G. Wu
Fermi National Accelerator Laboratory, Batavia, IL, USA
B. Guilfoyle, T. Reid, Argonne National Laboratory, Lemont, IL, USA

## Abstract

Electropolishing process and cathodes have undergone modification and optimization for both low- and high-beta 650 MHz five-cell niobium cavities for PIP-II. Cavities treated with these modified electropolishing conditions exhibited smooth surfaces and good performance in baseline tests. Nonetheless, due to administrative constraints on project cavities, maximum gradient performance testing was not conducted. This paper presents a study conducted on a single-cell 650 MHz cavity utilizing the optimized electropolishing conditions, highlighting the maximum performance attained for this specific cavity. The cavity tested at 2 K in a vertical cryostat reached a superior accelerating field gradient of 53.3 MV/m at $Q_0$ of $1.6 \times 10^{10}$, which is the highest gradient attained for this type of large-sized cavities.

## INTRODUCTION

Modern superconducting accelerator machines utilize niobium superconducting RF cavities, which operate in a superconducting state at 2 K. Various types of superconducting cavities are essential for building an efficient accelerating machine. Medium (~650 MHz) and high frequency (~1.3 GHz) elliptical cavities are used to accelerate high-energy particles. For an efficient accelerator, it is crucial to have cavities with both high gradient and high $Q_0$. SRF R&D focuses on enhancing these aspects, with surface processing playing a critical role in achieving optimal performance. Key processes include electropolishing, heat treatment, nitrogen doping (N-doping), mid-temperature baking, 120°C baking, and high-pressure water rinsing. Electropolishing is an electrochemical technique designed to remove interior material from the cavity, ensuring a defect-free and smooth surface.

1.3 GHz cavities have demonstrated excellent SRF performance with a high accelerating gradient of around 50 MV/m [1]. The corresponding surface peak magnetic field $B_{pk}$ reaches ~210 mT. 650 MHz niobium cavities should also be capable of reaching similarly high gradients and corresponding $B_{pk}$. However, the performance of 650 MHz cavities and cavities with similar frequencies reported so far have not met such a high performance.

Recently, we have demonstrated significant improvements in the surface quality and SRF performance of low- and high-beta five-cell 650 MHz cavities developed for the Proton Improvement Plan (PIP-II) linac [2, 3]. Notably, one of the HB650 cavities was tested beyond the administrative limit, achieving 29 MV/m without quenching [3]. The maximum performance potential of these cavities, following processing with the modified EP technique and upgraded cathode design, has yet to be fully assessed. This study focuses on applying the optimized EP conditions to a 650 MHz single-cell cavity with a $\beta$ value of 0.9 and evaluating its maximum quench field. This paper presents the surface processing methods applied and the resulting SRF performance of the 650 MHz single-cell cavity.

## EP SETUP AND CONDITIONS

The horizontal electropolishing (EP) setup and the cathode used in this study are illustrated in Fig. 1. The single-cell cavity (B9AS-AES-003) was processed using EP parameters detailed in Table 1. The cathode design used for the 5-cell cavities was scaled down and replicated for the cavity B9AS-AES-003. These EP parameters and cathode structure were consistent with those optimized for the 5-cell high-$\beta$ 650 PIP-II cavities [3].

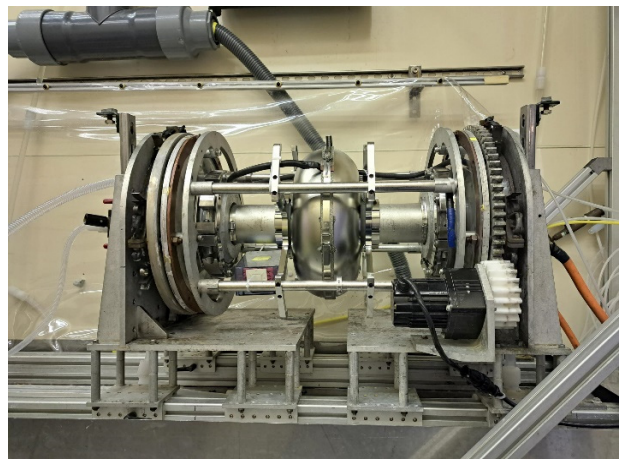

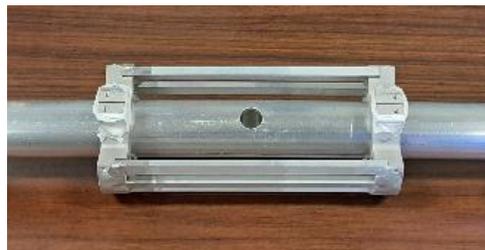

Figure 1: EP tool with the cavity B9AS-AES-003 (top). Patented cathode structure used for EP the cavity (bottom).

___________________
* This work was supported by the United States Department of Energy, Offices of High Energy Physics and Basic Energy Sciences under contract No. DE-AC02-07CH11359 with Fermi Research Alliance.
† vchouhan@fnal.gov

Table 1: Optimal EP Conditions Applied to the Cavity B9AS-AES-003

| Condition | Value/Detail |
| --- | --- |
| Cathode | New cathode with large surface area |
| Voltage | 20-25 V |
| Temperature | 12-22 °C |
| Electrolyte | Mixture of $H_2SO_4$ and HF |
| Cooling | Water spray on cavity |

## INITIAL CAVITY STATE

The cavity had been used in a previous study where it underwent EP with unoptimized conditions and cathode design, followed by nitrogen doping (N-doping). N-doping is a process employed to enhance the $Q_0$ of the cavity. An optical camera was used to inspect the equator surface, capturing images of the entire equator. The inspection revealed that the surface was rough and exhibited a defect, as shown in Fig. 2. Consequently, the SRF performance of the cavity was limited by quenching at an $E_{acc}$ of 16.6 MV/m. The $Q_0$ versus $E_{acc}$ curve is shown in Fig. 3.

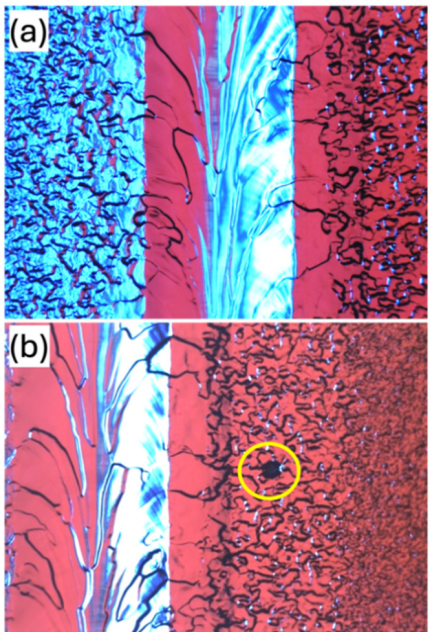

Figure 2: Optical images of the initial equator surface and the surface defect highlighted in (b). Image size: 12×9 mm.

## SURFACE PROCESSING

The cavity surface underwent several processing steps, including mechanical grinding, optimal EP, heat treatment, and low temperature baking. Since the cavity surface had the defect, the surface was repaired with local mechanical grinding. The grinding process used was described in detail elsewhere [4].

The cavity surface after the grinding process was treated with the optimal EP conditions and cathode structure. Two EP steps for 80 and 40 μm removal were applied with an intermediate vacuum furnace treatment at 900 °C for 3 h. Additionally, an in-situ vacuum bake at 120 °C for 48 h was applied.

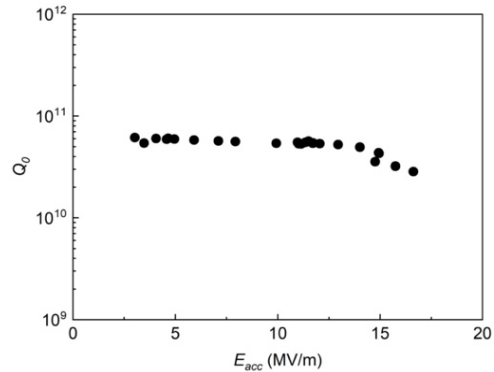

Figure 3: $Q_0$-$E_{acc}$ curve measured for the initial state of the cavity before treating with optimal EP.

## SURFACE AND REMOVAL

The cavity surface was inspected with an optical camera after the first EP step for 80 μm removal, but before the 900 °C treatment, to evaluate the impact of the optimized EP process. As shown in Fig. 4, the surface morphology at the equator position significantly improved, confirming that the EP conditions and cathode structure were optimal. Similar EP conditions have successfully produced smooth surfaces in multi-cell cavities as well [2-4].

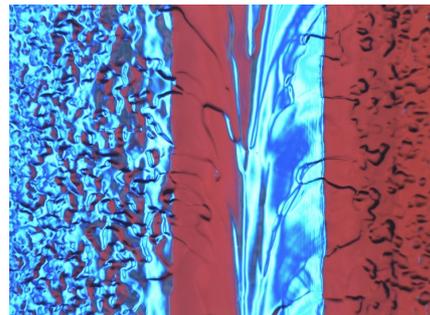

Figure 4: Optical image of the equator surface after optimal EP for 80 μm removal. Image size: 12×9 mm.

Using the large-sized cathode, which maintains close proximity to the irises, combined with a higher EP voltage, is typically considered risky to the potential for excessive material removal at the iris locations. Our previous study on multi-cell 650 MHz cavities has shown that the non-uniformity in material removal decreases with the use of similar large-sized cathode structure and higher EP voltage. However, removal near the iris was not measured due the inaccessibility of the ultrasonic thickness gauge probe between the cavity cells and stiffening rings near the irises.

To specifically study the effect of both the large cathode structure and higher voltage on material removal at the irises, cavity wall thickness was measured before and after

EP at the locations illustrated in Fig. 5. The removal thickness profile, shown in Fig. 5, was measured as close to the iris as possible to determine if the new EP conditions caused any excessive material removal.

The removal profile appears similar to what is typically observed for 1.3 GHz cavities electropolished with a cathode pipe that includes masking near the irises. The results indicate that even with a large cathode surface area and close proximity to the irises, coupled with a high applied voltage, there is no increased removal at the irises. The removal trend suggests that the EP conditions remain within safe limits, avoiding excessive material removal.

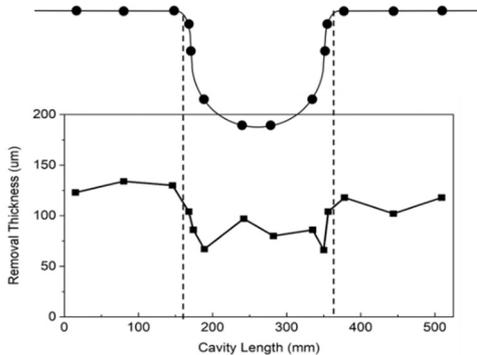

Figure 5: Removal profile for B9AS-AES-003 after being electropolished with the new cathode and optimal EP conditions.

## CAVITY PERFORMANCE

After the aforementioned surface processing, the cavity was tested in a vertical cryostat at 2 K. A second vertical test at 2 K was conducted after the cavity underwent an in-situ vacuum bake at 120 °C for 48 hours. The $Q_0$ versus $E_{acc}$ curves are presented in Fig. 6. In the first test, before the 120 °C bake, the cavity achieved an $E_{acc}$ of 35 MV/m, but its performance was power limited due to the presence of high field $Q$-slope (HFQS). After the standard 120 °C bake, the $Q$-slope was eliminated, and the cavity exhibited an exceptional accelerating gradient of 53.3 MV/m at a $Q_0$ of $1.6 \times 10^{10}$.

At 53.3 MV/m, the surface peak magnetic field $B_{pk}$ reached 198.2 mT, comparable to that seen in 1.3 GHz cavities which achieve high $E_{acc}$ above 45 MV/m. To the best of the authors' knowledge, the accelerating field gradient achieved for the 650 MHz cavity is the highest recorded to date. HFQS onset was estimated to be 108 mT which was similar to the typical onset value of ~100 mT in 1.3 GHz cavities [5]. These results demonstrate that our EP conditions and the designed cathode are optimal to deliver superior performance in 650 MHz cavities and those with similar frequencies.

The high performance of the cavity is attributed to the microscopic smoothness of the surface, achieved by smoothing sharp protrusions and grain boundary edges under the applied EP conditions. Although the surface roughness profile was not measured, the surface smoothening was apparent from the optical images.

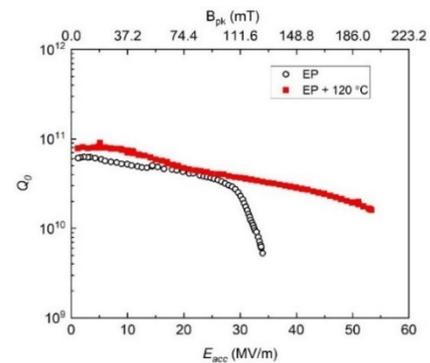

Figure 6: $Q_0$ versus $E_{acc}$ curves for 650 MHz single-cell cavity B9AS-AES-003 after optimal EP and 120 °C bake. $E_{acc}$ reached 53.3 MV/m, marking the highest performance recorded to date for 650 MHz cavities.

## CONCLUSION

Optimized EP conditions and a new cathode were applied to the 650 MHz single-cell Nb SRF cavity B9AS-AES-003. The cavity surface became smoother after the EP process. The modified EP conditions and the large-sized cathode structure were not found risky in causing a higher removal on the irises. The cavity achieved a superior accelerating gradient of 53.3 MV/m after EP and standard 120 °C bake, which is the highest gradient recorded for this type of large-sized cavity.

## ACKNOWLEDGMENTS

This manuscript has been authored by Fermi Research Alliance, LLC under Contract No. DE-AC02-07CH11359 with the U.S. Department of Energy, Office of Science, Office of High Energy Physics.

## REFERENCES

[1] D. Bafia, A. Grassellino, O. S. Melnychuk, A. S. Romanenko, Z-H. Sung, and J. Zasadzinski, "Gradients of 50 MV/m in TESLA Shaped Cavities via Modified Low Temperature Bake", in *Proc. SRF'19*, Dresden, Germany, Jun.-Jul. 2019, pp. 586-591. `doi:10.18429/JACoW-SRF2019-TUP061`

[2] V. Chouhan *et al.*, "Electropolishing parameters study for surface smoothening of low-β 650 MHz five-cell niobium superconducting radio frequency cavity," *Nucl. Instrum. Methods Phys. Res., Sect. A*, vol. 1051, p. 168234, Jun. 2023. `doi:10.1016/j.nima.2023.168234`

[3] V. Chouhan *et al.*, "Latest Development of Electropolishing Optimization for 650 MHz Cavity", in *Proc. SRF'23*, Grand Rapids, MI, USA, Jun. 2023, paper TUPTB042, pp. 512-516. `doi:10.18429/JACoW-SRF2023-TUPTB042`

[4] V. Chouhan, D.J. Bice, D.A. Burk, M.K. Ng, and G. Wu, "Visual, Optical and Replica Inspections: Surface Preparation of 650 MHz Nb Cavity for PIP-II Linac", in *Proc. 21th Int. Conf. RF Supercond. (SRF'23)*, Grand Rapids, MI, USA, Jun. 2023, pp. 507-511. `doi:10.18429/JACoW-SRF2023-TUPTB041`

[5] G. Ciovati, G. Myneni, F. Stevie, P. Maheshwari, and D. Griffis, "High field Qslope and the baking effect: Review of recent experimental results and new data on Nb heat treatments," *Phys. Rev. Spec. Top. Accel Beams*, vol. 13, no. 2, Feb. 2010. `doi:10.1103/physrevstab.13.022002`